\documentclass[twocolumn,secnumarabic,amssymb, nobibnotes, aps, showkeys]{revtex4-2}
\usepackage{amsmath}
\usepackage{mathtools}

\usepackage{algorithmic}
\usepackage{graphicx}
\usepackage{textcomp}
\usepackage{xcolor}
\usepackage{caption}
\usepackage{subfigure}

\setcitestyle{super}

\newcommand{\WSe}[1]{{$\mathrm{WSe_2}$#1}}
\newcommand{\eg}[1]{{\textit{e.g.,} #1}}

\newcommand{\new}[1]{{\color{black} #1}}
\newcommand{\newer}[1]{{\color{black} #1}}
\begin{document}
\title{Reduction of Magnetic Interaction Due to Clustering in Doped Transition-Metal Dichalcogenides: A Case Study of Mn, V, Fe-Doped $\rm WSe_2$}
\author{Sabyasachi Tiwari$^{1,2,3*}$, Maarten~Van~de~Put$^{1,2}$, Bart~Sor\'ee$^{2}$$^{4}$$^{5}$, Christopher Hinkle$^{6}$ and William~G.~Vandenberghe$^{1}$}
\email{william.vandenberghe@utdallas.edu}
\email[$^*$]{sabyasachi.tiwari@austin.utexas.edu;}
\address{$^1$ Department of Materials Science and Engineering, The University of Texas at Dallas, 800 W Campbell Rd., Richardson, Texas 75080, USA.}
\address{$^2$ Imec, Kapeldreef 75, 3001 Heverlee, Belgium.}%
\address{$^3$Department of Materials Engineering, KU Leuven, Kasteelpark Arenberg 44, 3001 Leuven, Belgium}
\address{$^4$Department of Electrical Engineering, KU Leuven, Kasteelpark Arenberg 10, 3001 Leuven, Belgium.}%
\address{$^5$Department of Physics, Universiteit Antwerpen, Groenenborgerlaan 171, 2020 Antwerp, Belgium.}
\address{$^6$Department of Electrical Engineering, University of Notre Dame, Notre Dame, IN 46556, USA.}

\begin{abstract}
\textbf{Abstract: \new{Using Hubbard U corrected density functional theory calculations, lattice Monte-Carlo, and spin-Monte-Carlo simulations, we investigate the impact of dopant clustering on the magnetic properties of \WSe~doped with period four transition metals.}
We use manganese (Mn) and iron (Fe) as candidate n-type dopants and vanadium (V) as the candidate p-type dopants, substituting the tungsten (W) atom in \WSe.
Specifically, we determine the strength of the exchange interaction in the Fe-, Mn-, and V-doped \WSe~ in the presence of clustering.
We show that the clusters of dopants are energetically more stable than discretely doped systems.
Further, we show that in the presence of dopant clustering, the magnetic exchange interaction significantly reduces because the magnetic order in clustered \WSe~becomes more itinerant.
Finally, we show that the clustering of the dopant atoms has a detrimental effect on the magnetic interaction, and to obtain an optimal Curie temperature, it is important to control the distribution of the dopant atoms.}
\end{abstract}


\maketitle

\section{Introduction}
Two-dimensional (2D) magnetic materials like $\rm CrI_3$~\cite{sample11,sample24,sample19}, $\rm CrGeTe_3$~\cite{sample20},  and $\rm VSe_2$\cite{VSe2} have sparked great interest for their possible use in applications including, spintronics\cite{sample21}, valleytronics~\cite{sample37}, the realization of skyrmions~\cite{sample22},  magnetic memories,  and topological phase transition based devices~\cite{sample23}.
Many new experimental~\cite{Tiancheng2018, Wahab2021,Xu2022,Xie2022}, as well as theoretical ~\cite{Tiwari2023},  avenues are being explored to realize new applications of 2D magnetic materials.

However, the magnetic order in 2D magnetic crystals suffers due to their low magnetic anisotropy and weak exchange interaction strength~\cite{my_paper}, resulting in a low Curie temperature (e.g. 45 K for $\rm CrI_3$ and 42 K for $\rm CrGeTe_3$) which limits their practical application.
The prospects of 2D magnetic crystals for practical applications are also marred by the thermal instability of their magnetic order even below the Curie temperature~\cite{my_paper}.

One avenue that is being explored to circumvent the low Curie temperature of crystal 2D magnets is to dope conventional semiconducting 2D materials with an impurity atom.
In this regard, metal-doped transition-metal dichalcogenides (TMDs) have attracted major attention~\cite{sample14,sample8,sample30,sample36,Tiwari2021npj,V_dope,sample31,sample32,Reyntjens_2021,Reyntjens_2020}.
Doped semiconductors couple the properties of semiconductors and magnets and are called dilute magnetic semiconductors (DMS).
The ability to control the magnetic order through charge transfer in DMSs~\cite{DMS_1,DMS_2} also opens up the possibility of realizing magnetic devices because their magnetic state can be controlled using an external electric field~\cite{spin_tran_1}.

\begin{figure*}[t]
	\centering  
   \includegraphics[width=1.9\columnwidth]{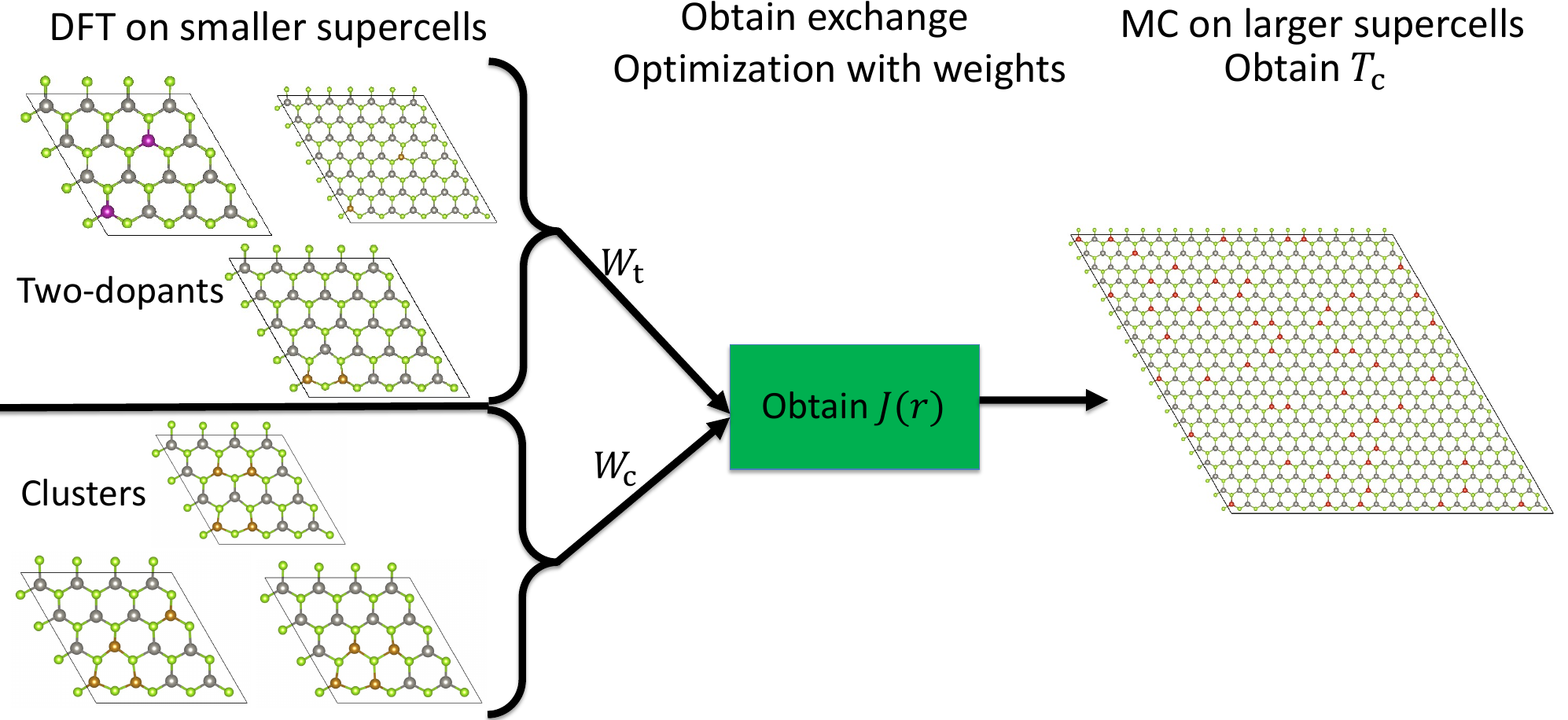}
    \caption{Our model for calculating the Curie temperature with the dopant concentration (Mn, Fe and V) in \WSe~from DFT and MC.  In the first step we perform DFT calcualtions with constrained magnetic orientation for two-dopant systems and clustered systems. We next statistically weight the doped systems to obtain the exchange interaction $J(r)$ using Eq.~(\ref{e:objectivefunction}). Finally, we obtain the phase transition using Monte-Carlo algorithm applied on a larger supercell of \WSe~doped with certain doping concentration.}
	\label{f:scheme}
\end{figure*} 

Although there have been many theoretical predictions of robust ferromagnetism in metal-doped TMDs~\cite{sample14,sample8,sample30,sample36,Tiwari2021npj}, an experimental realization has remained elusive.
There have been many recent experimental works that report magnetic order in metal-doped TMDs~\cite{kochat2017,V_dope, Yun2021,sample31,sample32,Tao2022,Zhang2023,Pathirage2023,Zhang2020} but there are many inconsistencies in their properties.

For example, in Ref.~\cite{kochat2017} the authors observe a magnetic order at a higher doping concentration and a low temperature in Re-doped $\rm MoTe_2$, but the magnetic order appeared to be a superposition of paramagnetic and ferromagnetic states, suggesting a spin-glass phase.
On the contrary, in Ref.~\cite{V_dope} the authors dope V in \WSe~and observe room-temperature ferromagnetism at a 0.1\% doping concentration.
\newer{However, in their measurements, they observe the effect of very small clusters of V dopants.}
As soon as they increase the dopant concentration, their Curie temperature starts decreasing which they attribute to multiple domain formation at high doping concentrations with opposite or random spin configurations.
Other works including Ref.~\cite{sample31,sample32,Tao2022} have measured the hysteresis curves in Ni-doped \WSe,~Ta-doped $\rm MoTe_2$, and Fe-doped and V-doped $\rm MoS_2$ at room temperature.
However, the presence of hysteresis curves at room temperature can be due to stray magnetism and may not confirm a fully ferromagnetic sample~\cite{V_dope}.

{There have been many theoretical works that have attempted to understand the magnetic order in doped TMDs~\cite{sample14,sample8,sample30,sample36,Tiwari2021npj,Shu2015}.}
Previous theoretical predictions of the Curie temperature of doped TMDs have predicted Curie temperatures $>1000$ K~\cite{sample14,sample8,sample30,sample36}, whereas experimental observations to date suggest a Curie temperature $<350$ K~\cite{sample31,sample32}, which varies sample to sample~\cite{V_dope}.
The theoretical works predicting $> 1000$ K Curie temperatures have ignored the effect of magnetic anisotropy~\cite{sample27} and have used the collinear magnetic approximation~\cite{J1_J2,sample8,sample14}.
More refined theoretical calculations have predicted a much lower and much richer magnetic order in doped TMDs~\cite{Tiwari2021npj}.
The theoretical Curie temperature calculations report the bulk Curie temperature, whereas, in experiments, the mere existence of a spin-polarization or hysteresis curves at a certain temperature is reported as the Curie temperature~\cite{V_dope}.
Nevertheless, the source of experimental inconsistencies and ill-match to theoretical predictions in doped TMDs remains an open question.

A major effect that is ignored in most theoretical predictions of magnetic order in TMDs is the effect of dopant clustering.
Clustering of dopants is quite dominant in the three-dimensional DMSs, \eg~Mn-doped GaAs and GaN\cite{clus_zunger,sample13, sample33, sample34}, and decides the upper limit of the dopant concentration that could lead to a finite Curie temperature~\cite{clus_zunger,sample13}.

In this work, we investigate the effect of the clustering of dopant atoms on the magnetic structure of \WSe.
We choose \WSe~as the base TMD doped with Mn/V/Fe due to its high magnetic anisotropy~\cite{Tiwari2021npj}.
Moreover, there are recent experimental realizations of magnetic order in \WSe~\cite{sample31,V_dope, Yun2021,Pathirage2023} motivating our material choice.
First, we briefly describe the anisotropic Heisenberg model and the exchange interaction function ($J(r)$) used to approximate the magnetic interaction in doped \WSe, and our method of calculating formation energies for dopant configurations in TMDs.
We next show the energy of formation as a function of two dopant separations in TMDs and the magnetic exchange as a function of dopant distance.
Finally, we discuss the effect of the clustering of the dopant atoms on the magnetic exchange interactions in doped \WSe~ and show that the clustering of dopants negatively impacts the magnetic order in \WSe~which can vary from sample to sample.

\section{Model}

We model the magnetic structure using the classical Heisenberg Hamiltonian,
\begin{equation}
    H
    = -\sum_{i,j} \mathbf{S}_{i}J_{ij}\mathbf{S}_{j}+D\sum_{i}(\hat{S}_{i}^z)^2.
    \label{e:hamiltonian}
\end{equation}
The first term is the exchange term between the $i^{\rm th}$ and the $j^{\rm th}$ magnetic atom (dopant) with $\mathbf{S}=S_x\mathbf{x}+S_y\mathbf{y}+S_z\mathbf{z}$, as the magnetic moment vector.
$J_{ij}$ is the strength of the exchange interaction between the $i^{\rm th}$ and the $j^{\rm th}$ magnetic atom and is a tensor as described in Ref. \cite{my_paper}.
We only use the diagonal elements of $J_{ij}$ which are, $J_{ij}^{xx}$, $J_{ij}^{yy}$, $J_{ij}^{zz}$, which measure the strength of magnetic exchange interaction for the magnetic axis in $x$, $y$, and $z$ direction, respectively.
Because of the in-plane isotropy in TMDs,  $J_{ij}^{xx}=J_{ij}^{yy}=J_{ij}^{\parallel}$ and $J_{ij}^{zz}=J_{ij}^{\perp}$, with $J^\parallel$ being the in-plane exchange interaction, and $J^{\perp}$ being the out-of-plane exchange interaction.

%The second term in Eq. (\ref{e:hamiltonian}) is the onsite anisotropy term, with $D$ being the strength of the onsite anisotropy.
There are two types of anisotropies, the onsite anisotropy, whose strength is quantified by $D$ in Eq. (\ref{f:Curie}), and the exchange anisotropy, whose strength is quantified by $\Delta_{\mathrm{ex}}=J^{\perp}-J^{\parallel}$.
We follow the procedure outlined in our previous work~\cite{Tiwari2021npj} to obtain the exchange interaction tensor ($J_{ij}$) and the onsite anisotropy ($D$).

The elements ($J^{\perp/\parallel}$) of the interaction tensor ($J_{i,j}$) are approximated using a functional form,
\begin{multline}
    J^{\perp/\parallel}(r_{ij}) = c^{\perp/\parallel}\sum^{3}_{i=1}c_{i}B_i(r)h(r_{\mathrm{c}}-r)+\\
      c\exp(-r/\lambda)h(r-r_{\mathrm{c}}).
    \label{e:J}
\end{multline}
Here, $h(r)$ is the Heaviside step function. 
$r_c$ is the cut-off radius within which we approximate the $J$ parameters using a sum of spline functions $B_i(r)$, and outside $r_c$ we approximate $J(r)$ using the exponential $c\exp(-r/\lambda)$.
\new{The parameters $c^{\perp/\parallel}$, $c_i$ are obtained from the DFT using our method described in Ref.~\cite{Tiwari2021npj}.
$c$ and $\lambda$ are obtained by matching the spline functions and their derivatives with the exponential function and its derivative at $r=r_{\rm c}$.  
The spline functions allow us to statistically average the strength of exchange interaction due to clustering, and to account for the long-range interaction more efficiently.}

\new{As shown in Fig.~\ref{f:scheme}, we obtain the $J(r)$ from smaller supercells, $4\times4\times1$,  $5\times5\times1$, and $7\times7\times1$ of doped \WSe~using DFT \cite{my_paper,Tiwari2021npj}.
We then simulate multiple bigger supercells of size $30\times30\times 1$ using MC simulations to obtain the median and the variance of the critical temperature as a function of dopant concentration.}

\new{We calculate the energy of formation using,
\begin{equation}
	E_{\rm form}=\big(E_D-N_SE_B-\Delta N_T \mu_T-\Delta N_X \mu_X- \Delta N_M \mu_M\big).
\label{e:form1}
\end{equation}
Here $\Delta$ represents the change in the number of atoms after doping for respective material and $E_D$ is the total energy of the doped system.
$N_T$ is the number of transition metal, $N_X$,  is the number of chalcogen atoms, and $N_M$ is the number of metal dopants. 
In our case, $T: \rm W$, $X: \rm Se$, and $M: \rm Fe,\,V,\,Mn$.
$E_B=\mu_{\rm T}+2\mu_{\rm X}$ is the total energy of the bulk TMD and $N_S$ is the supercell size.
$\mu$ are the chemical potentials.
Substituting these in Eq.  (\ref{e:form1}), we obtain the energy of formation per dopant atom as a function of $\mu_{\rm Se}$,
\begin{align}
	%E_f=N_SE_b-(N_T-N_M)\frac{N_SE_b-2N_T\mu_X}{N_T}-2N_T\mu_X-N_M\mu_M
	E_f &=\frac{1}{\Delta N_M}\big(E_D-{N_S}E_B -{\Delta N_{\rm W}}E_B- \nonumber\\
	      &{(\Delta N_{\rm Se}-2\Delta N_{\rm W} )}\mu_{\rm Se} -{\Delta N_M} \mu_M\big)
\label{e:form2}
\end{align}
}
\new{For all calculations reported below, we sweep the chemical potential of Se from its Se-rich to Se-deprived phase.
The chemical potential for the Se-rich condition is obtained by calculating the total energy of the Se atom in its crystalline phase and turns out to be $\mu_{\rm Se}$ (rich)=-3.12 eV.
We obtain the chemical potential of the Se-deprived state using $\mu_{\rm Se}= (E_{\rm B}-\mu_{\rm W})/2$. Where $\mu_{\rm W}$ is the chemical potential of a tungsten atom in its bulk phase, which turns out to be $\mu_{\rm W}=-10.44$ eV.
The chemical potentials used for Fe, Mn, and V are -7.99, -9.126, and -9.470 eV, respectively.
All DFT total energy calculations for obtaining the chemical potentials are performed with spin-orbit-coupling turned on.}
%From the obtained $J$ parameters, we calculate the Curie temperature by simulating the magnetic phase change of the Heisenberg Hamiltonian of equation~\ref{e:hamiltonian} using the MC simulations and taking the maximum of susceptibility ($T_{\mathrm{C}}=\mathrm{argmax}(\chi(T))$, where $\chi$ is the susceptibility)\textbf{ref my paper}.

\section{Results and Discussion}
\begin{figure}[th]
  \centering
  \subfigure[]{\includegraphics[width=0.49\columnwidth]{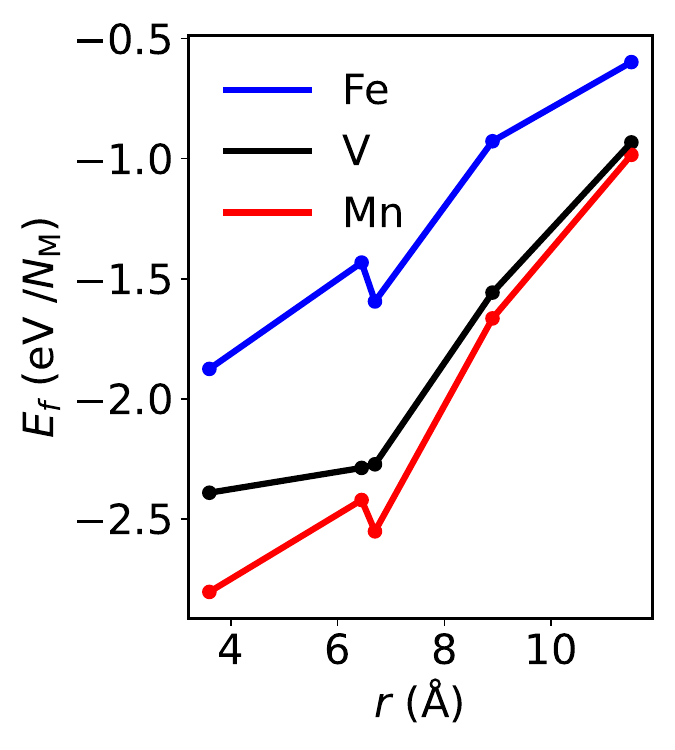}}   
  \subfigure[]{\includegraphics[width=0.49\columnwidth]{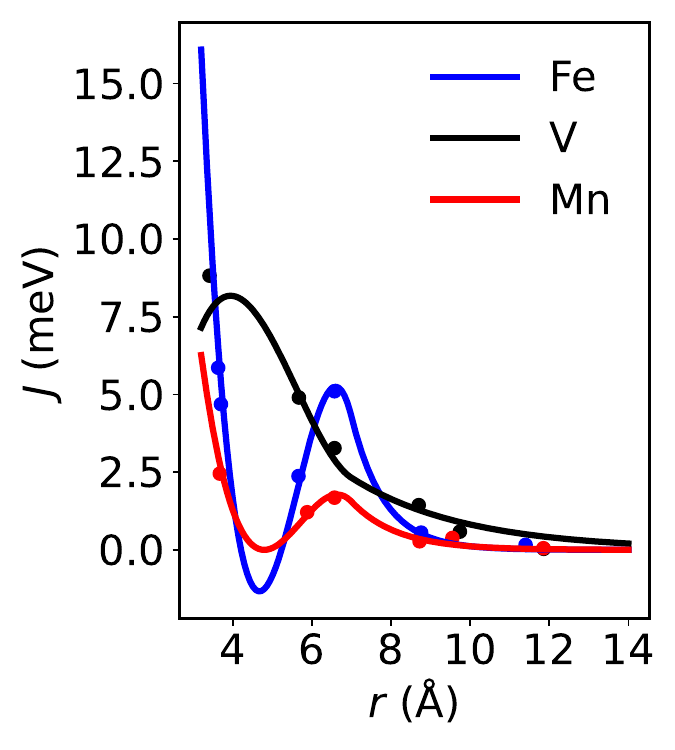}}
%  \subfigure[]{\includegraphics[width=0.7\columnwidth]{bar_plt_D.pdf}}
  %\subfigure[]{\includegraphics[width=0.7\columnwidth]{bar_plt_M.pdf}}
  \caption{(a) The energy of formation as a function per dopant ($E_f$) as a function of dopant distance for Fe, V, and Mn dopants in \WSe.  (b) The magnetic exchange interaction $J(r)$ as a function of dopant distance. }% 
  \label{f:J_plt} 
\end{figure}

\begin{figure*}[th]
  \centering
 {\includegraphics[width=1.8\columnwidth]{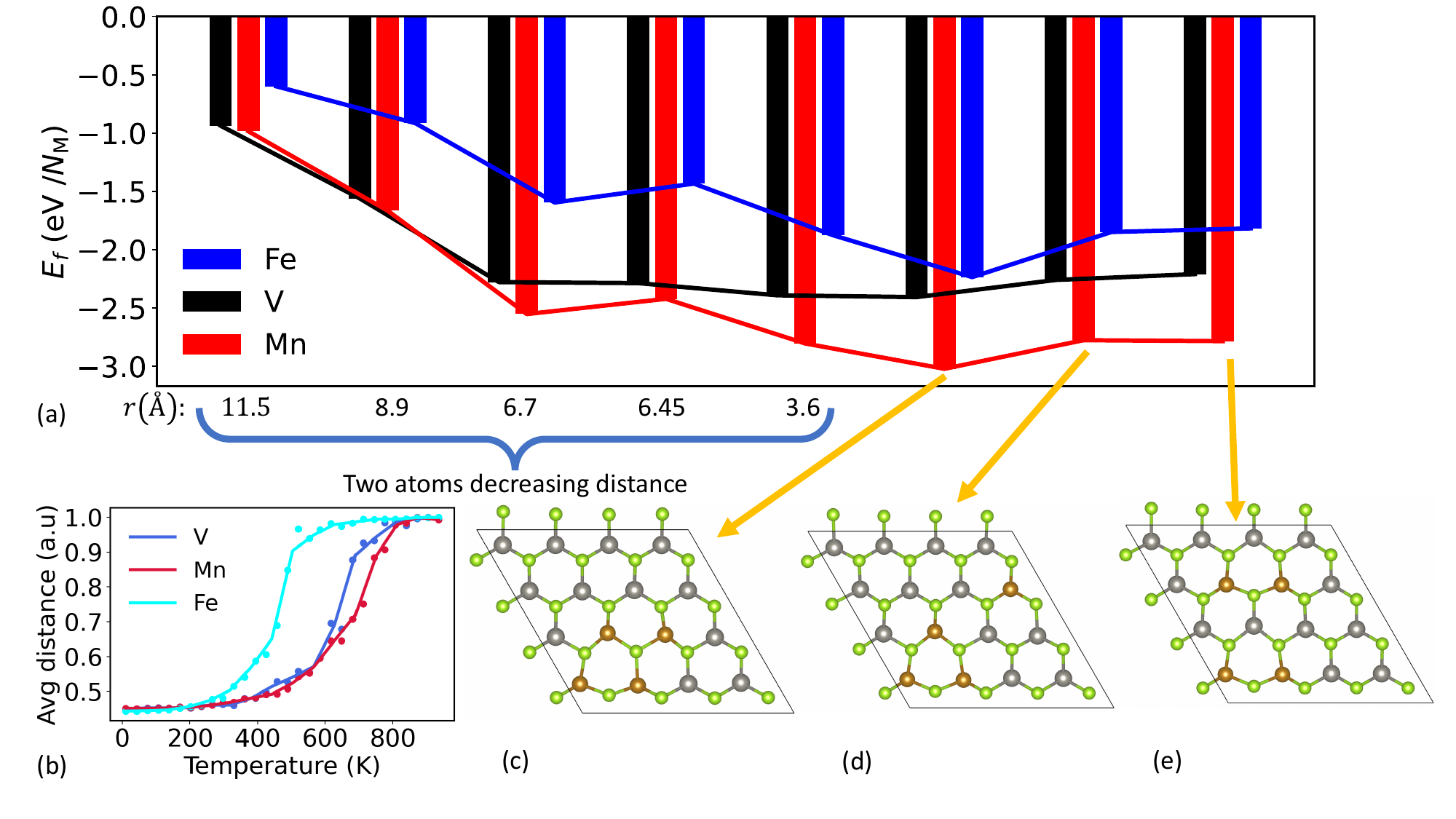}}
 \caption{(a) The formation energies per dopant atom ($N_{\rm M}$) $E_f$ of Fe-, V-, and Mn-doped \WSe~ for two substitutes (1-5) whose separation length  ($r (\rm \AA)$) is shown in the bottom,  and for four substitutes in clustered configurations (6-8) shown in (c-e).  The lower energy of formation means more stablity. \new{(b) The average dopant distance obtained using the lattice Monte-Carlo simulations for Fe-,V-, and Mn-dopants  (at 10\% doping concentration) as a function of temperature.  The solid line is a guide to the eye fit to the dotted MC data.}  } 
  \label{f:formation_cluster} 
\end{figure*}

%We choose \WSe~as the base TMD doped with Mn/V/Fe due to it's high magnetic anisotropic energy~\cite{Tiwari2021npj}. 
%Morever, there are recent experimental realizations of magnetic order in \WSe~\cite{sample31,V_dope} leading to our choice.
%It has been shown previously that magnetic anisotropy is the most important parameter when it comes to the Critical temeprature~\cite{approach_1}. 

\subsection{Magnetic exchange parameters of unclustered \WSe~}

\begin{figure}[th]
  \centering
	{\includegraphics[width=0.9\columnwidth]{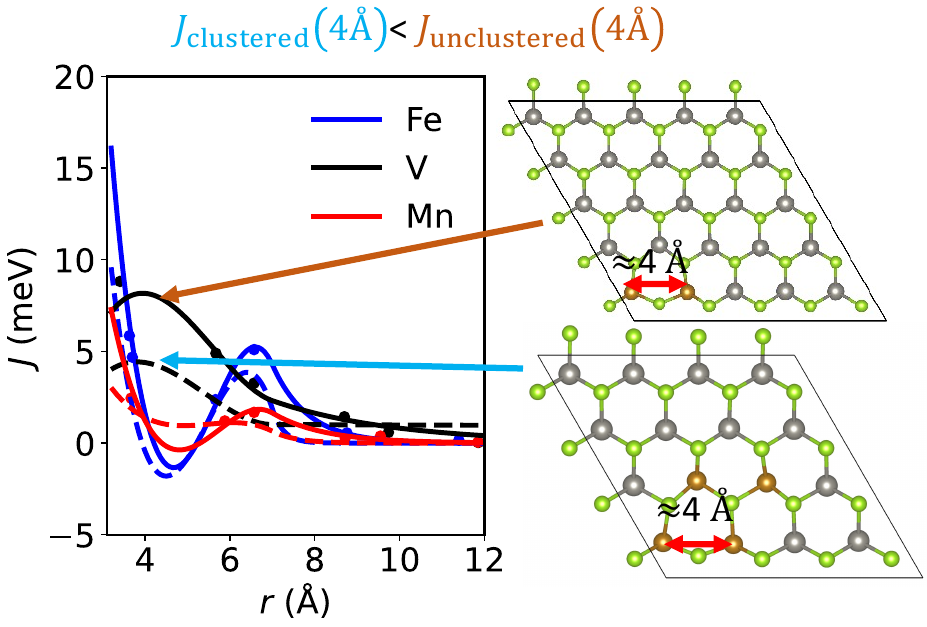}}
 \caption{{The average magnetic exchange interaction $J(r)=(J_\parallel (r)+J_\perp (r))/2$ as a function of dopant distance (left) for Fe, V, and Mn dopants in the presence of clustering (dotted lines) and in the absence of clustering (solid lines). } We have chosen a clustering of $\alpha_k=10\%$ (see Sec. ~\ref{s:method} for details). The dots show the calculated exchange parameters using the $J_1-J_2$ method in the absence of clustering.  {Top and bottom right figures depict a supercell of doped \WSe~ in the absence (top) and presence (bottom) of clustering. The clustering of dopants (shown right bottom) reduces the overall exchange interaction.}} 
  \label{f:clustering} 
\end{figure}

\new{In Fig.~\ref{f:J_plt} (a), we plot the energy of formation per unit area {per dopant} as a function of dopant separation distance $(r)$.}
Here we only plot for Se-rich conditions, under which all structures are energetically stable.
We observe that with decreasing distance, the energy of formation becomes more negative, suggesting that the closer the dopants, the more stable the structure is.
The formation energy for the structures with dopant atoms at a distance $r\approx 3.2 \AA$ is almost 200\%, 134\%, and 200\% more stable than $r\approx 11 \AA$ for Fe, V, and Mn dopants, respectively.

In Fig.~\ref{f:J_plt} (b), we plot the $J(r)$ (solid lines), which is the combination of spline functions ($r_{\mathrm{c}}<7$ \AA) and the exponential screening ($r_{\mathrm{c}}\geq7$ \AA), and the discrete $J$ parameters (dots) obtained by dividing the energy difference between the FM and the AFM configuration by the square of magnetizations of the dopant atoms ($M$).
We find that our chosen function to approximate $J(r)$ is a good fit for the discrete $J$ values.

From Fig.~\ref{f:J_plt} (b), we observe that the exchange interaction ($J(r)$) is the strongest for Fe and V as dopants whereas Mn has the lowest $J(r)$.
The long-range interaction in V is the strongest, and the short-range interaction (up to 7 \AA) is the strongest in Fe.
For $r<4 \AA$, the exchange interaction is as high as 15, 8.0, and 6.0 meV for Fe, V, and Mn dopants, respectively.
These exchange interactions are higher than most of the 2D crystals where the maximum is found for $\rm CrGeTe_3$ of 2.8 meV~\cite{my_paper}.
The strong $J(r)$ would lead to a higher critical temperature when the dopants are close to each other.
However, if the dopants start forming too close, then the clusters would be separated by a long range.
This would in turn lead to a lower critical temperature because the long-range exchange interaction is weaker for all three dopants and decays exponentially.

\begin{figure}[ht]
	\centering  
   \includegraphics[width=0.99\columnwidth]{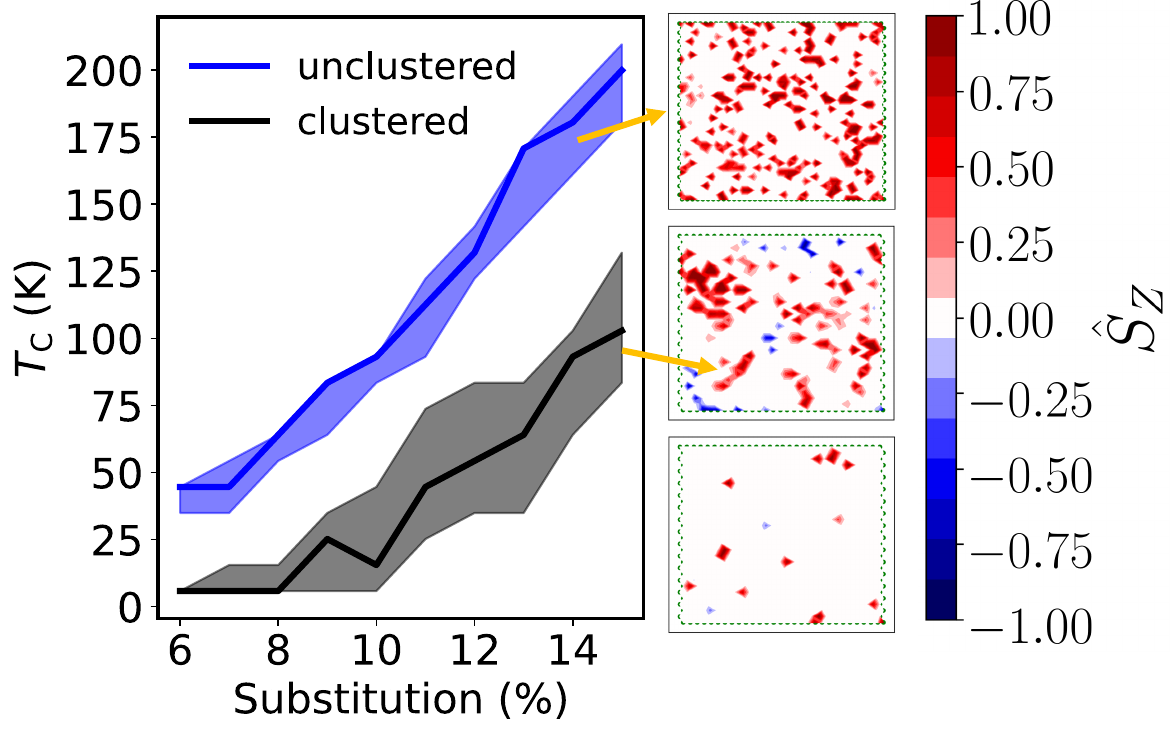}
    \caption{The Curie temperature of V-doped \WSe~as a function of dopant substitution in the absence (blue) and presence (black) of clustering. The filled region shows the 25th to 75th percentile of the variation of Curie temperature accross samples. The contour plots in the right show the magnetic order in a sample of \WSe~obtained from the MC simulation for unclustered (top) and clustered (middle) at a dopant substitution of 15\% and temperature of 93 K.  The bottom contour plot shows the magnetic order of a sample at a doping of 1\% and a magnetic field of $0.1 \rm meV/\mu_B$ at 93 K. At 1\% all spins align to the applied magnetic field but there is no ferromagnetic order.}
	\label{f:Curie}
\end{figure} 

\begin{figure*}[th]
  \centering
%  \subfigure[]{\includegraphics[width=1.8\columnwidth]{clustr.pdf}}\\
  \subfigure[]{\includegraphics[width=0.66\columnwidth]{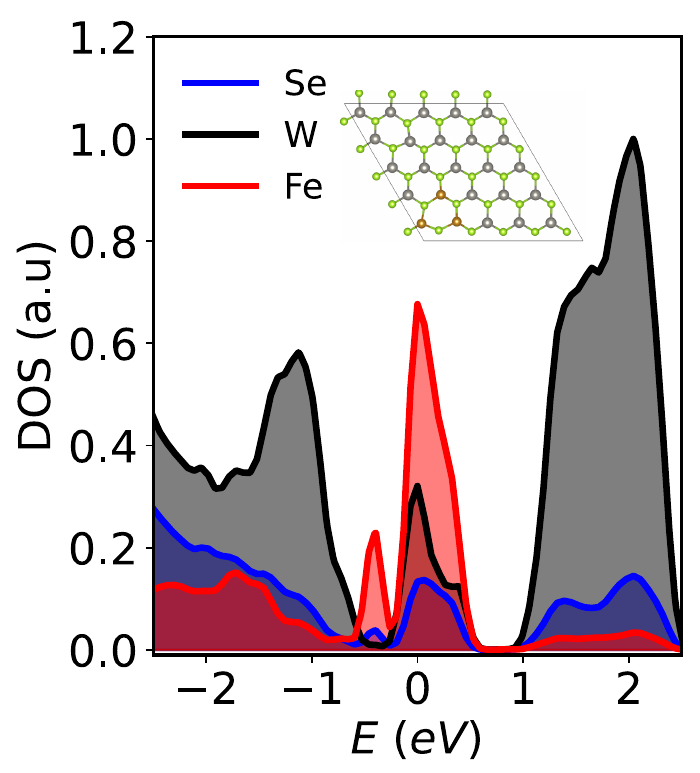}}
  \subfigure[]{\includegraphics[width=0.66\columnwidth]{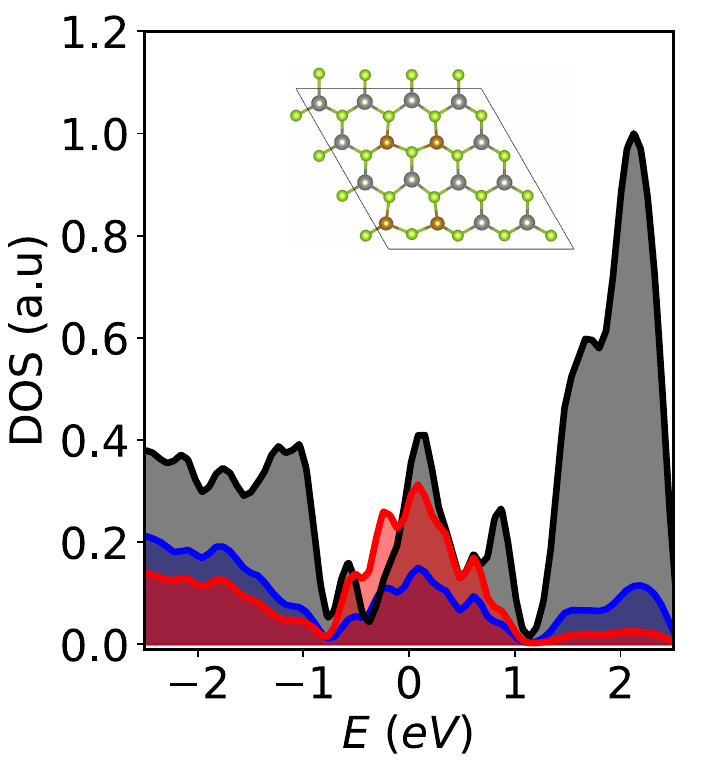}}
  \subfigure[]{\includegraphics[width=0.64\columnwidth]{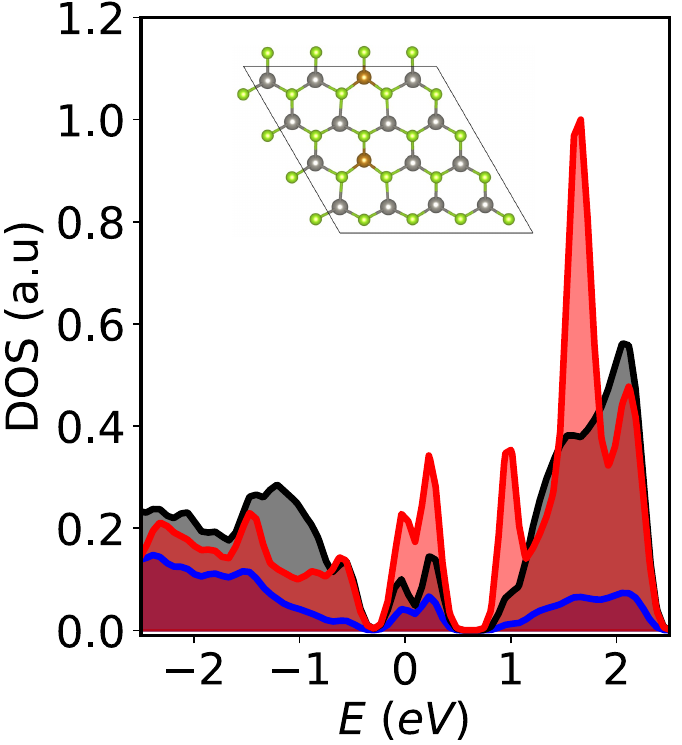}} 
 \caption{The atomic type resolved density of states for the supercells shown in the inset \newer{with doping concentrations (W-substitution) (a) 12\%,  (b) 25\%,  and (c) 12.5\%}.  The density of states is calculated per atom.  } 
  \label{f:dos} 
\end{figure*}

%\begin{figure*}[th]
  %\centering
  %{\includegraphics[width=\columnwidth]{1.pdf}}
 %\caption{(a) Shows the supercells of \WSe clustered with dopants. (b) Shows the  $J(r)$ for the different levels of clustering (solid lines) and the $J$ parameters (dots) for the individual structures (both clustered (yellow and pink) and unclustered (blue and green)).} 
%  \label{f:clustering2} 
%\end{figure*}
\subsection{Impact of clustering} 

\new{Figure~\ref{f:formation_cluster} (a) shows the energy of formation per unit area $E_f$ for Fe, V, and Mn dopants in \WSe~for two substitutes as well as for clusters of dopants shown in Fig. ~\ref{f:formation_cluster} (c-e).
We call a structure clustered if the supercell has more than two dopant atoms and the root-mean-square (RMS) distance between dopants is below $7$ \AA.
We observe that the clustered structures have lower energy of formation, indicating that they are energetically more stable than the unclustered structures.
We also observe that $E_f$ is much lower for clustered structures for Fe and Mn dopants in \WSe~ compared to V dopants.
This suggests that even at lower doping concentrations, the likelihood of clustering is significantly higher in Fe- and Mn-doped \WSe.
It should be noted that the formation energies in Fig.~(\ref{f:formation_cluster}) are per dopant atoms, and for a comparison of cluster stability with temperature, the number of dopants and their configurations need to be taken into account.
}

{
Using lattice Monte-Carlo simulations \cite{LMC}, where energy is parameterized using the energy of formation shown in Fig. (2), we find that clusters of Fe-, V-, and Mn-dopants can be stable up to 400, 600, and 700 K at a 10\% doping as shown in Fig.~\ref{f:formation_cluster} (b) (see supplementary for more information~\cite{SI}).
Interestingly, we see that there is evidence of a first-order phase transition from clustered to liquid (random position of dopants) phase in doped \WSe.
}

We now discuss the impact of dopant clustering on the $J$ parameters {obtained from the DFT} in doped \WSe.
As discussed in the methods section, the clustering of dopants significantly alters the exchange interaction strength.
Moreover, the clustering of dopants also breaks the rotational symmetry, and hence, the obtained $J(r)$ is an ensemble average of the exact $J$ that one obtains for various clustering configurations.
Here, for ensemble averaging, we assign a 10\% weight to the clustered structures (more information in the methodology section~\ref{s:method}).

Figure ~\ref{f:clustering} shows the obtained $J (r)$ for clustered (dashed lines), unclustered (solid lines), and discrete $J$ parameters (dots) for \WSe~doped with Fe, V, and Mn dopants.
We observe that the exchange interaction reduces with the clustering of the dopant atoms.
Similar to Mn, both Fe- and V-doped \WSe~show lowering exchange interaction with the clustering of dopants.
Therefore, the clustering of dopant atoms in a sample of \WSe~is detrimental to the critical temperature of doped \WSe.

%We now present the calculated Curie temperture as a function of dopant concentration for Fe/Mn/V doped \WSe.

Figure~\ref{f:Curie} shows the variation of Curie temperature as a function of dopant substitution for V-doped \WSe~which is calculated using the $J(r)$ obtained for clustered and unclustered systems.% (For Fe- and Mn-doped \WSe, refer to the supplementary document).
The solid line shows the median of the critical temperature obtained for a particular doping concentration, whereas the shaded region shows the $25^{\rm th}$ to $75^{\rm th}$ percentile of the Curie temperature.
We define the doping concentration using, $\frac{N_{\mathrm{Dopant}}}{N_{\mathrm{W}}+N_{\mathrm{Dopant}}}\times 100$, where $N_{\rm Dopant}$ is the number of dopant atoms and $N_{\rm W}$ is the total number of $\rm W$ atoms in the \WSe~supercell.
At 15\% doping, $T_{\rm c}$ drops from 200 K to 100 K due to clustering while at 10\% doping, Curie temperature drops from 100 K to 25 K.
Below 10\% doping, no ferromagnetism is observed for clustered structures.
Therefore, the drop in $J(r)$ due to clustering, almost halves the Curie temperature.

In Fig. \ref{f:Curie}, the contour plots on the right show the magnetic order in a sample of \WSe~obtained from the MC simulation for unclustered (top) and clustered (middle) samples at a dopant substitution of 15\% and a temperature of 93 K.
We observe that in the case of unclustered samples, the magnetic order is strongly ferromagnetic, while in the case of clustered samples, there are magnetic domains with opposite magnetization that form.
{For the clustered system, we see two effects taking place.
The first effect is that the clusters are localized, and cluster-cluster interaction is long-range.
Weaker inter-cluster interaction works against a uniform ferromagnetic order across samples.
The second effect is that when the clusters form, their short-range exchange interaction also reduces as shown in Fig.~\ref{f:clustering}, which leads to a weaker intra-cluster exchange interaction.
These two effects combined lead to an overall weakly ferromagnetic/paramagnetic order.
}

The effect of clustering will result in an optimal doping concentration to realize the highest possible Curie temperature.
{Going to higher doping concentrations will increase clustering, and clustering can reduce Curie temperature or even lead to paramagnetic order.}
Hence, while doping, it is necessary to control the diffusion of dopants because even at low doping concentration, dopants can form clusters~\cite{clus_zunger} and reduce the effective exchange interaction strength leading to a reduced critical temperature.

Interestingly, in Fig.~\ref{f:Curie}, the bottom contour plot is for a sample with just 1\% doping and a small magnetic field of $0.1 \rm meV/\mu_B$.
We can see that the V-dopants orient in a majority spin-up orientation.
This is exactly what is reported in many experimental works that confirm the presence of magnetic order by using magnetic force microscopy (MFM)~\cite{V_dope,Yun2021}.
{However, it can be seen that a uniform magnetic order in the entire sample is missing.
Therefore, the MFM experiments of magnetic order at low doping densities in TMDs confirm the localized nature of magnetic order.
Unfortunately, as we see from our theoretical calculations, a sample-wide ferromagnetic order may still be missing even if the dopants form localized magnetic order. A higher doping induces clustering which would further hinder a uniform ferromagnetic order.}

{To further analyze the reason for lower exchange interaction strengths, we show the density of states resolved by atomic type in Fig.~\ref{f:dos} for Fe-doped \WSe.
From (a) to (c) the level of clustering decreases, however, the dopant concentration is higher in (c, 12.5 \%) compared to (a, 12~\%).

\new{We observe that both in (a) and (b) (higher clustering cases), the localized defect states around the Fermi level hybridize with the valence and conduction bands of the host material, resulting in an itinerant/delocalized magnetic behavior.}
While, in the case of (c), the defect states do not hybridize with the valence and conduction bands maintaining the localized nature.
The itinerant behavior reduces the localized nature of the magnetization on the dopant atoms, resulting in a weakened exchange interaction because the electrons can delocalize to reduce their kinetic energy.}

\section{Conclusion}

We have theoretically investigated the impact of dopant cluster formation on the magnetic order of Fe, V, and Mn-doped \WSe.
The most significant result of our work is that the clustering of dopants is favored energetically in \WSe.
\new{We further performed lattice MC simulations to determine the temperature up to which dopant clustering can happen in \WSe.
We found that at about 10\% doping V- and Mn-doped \WSe~can cluster up to 700-800 K (see supplementary for results on 14\% clustering~\cite{SI}).
It should be noted that our lattice MC simulations are quite simplistic as they do not account for charge migration and for precise estimation one has to perform ab-initio molecular dynamics simulations (AIMD)~\cite{Rev1a,Rev1b}.}
\new{We quantify the temperatures up to which clusters in TMDs can remain stable using lattice-switch Monte-Carlo calculations. We note that a quantitatively more accurate technique is to use ab-initio molecular dynamics (AIMD) to capture intersite potential barriers. However, in our case, as we deal with large supercells and higher doping concentrations, it is impossible to perform AIMD calculations due to excessive memory requirements.
}

The major impact of clustering is that clustering is significantly detrimental to the strength of the magnetic exchange interaction.
Therefore, clustering is detrimental to realizing magnetic order in \WSe.
We found that the lowering of the exchange interaction due to dopant cluster forming is a result of a transition from localized to itinerant magnetic order in \WSe.
Therefore, at higher doping concentrations, such a transition would be inevitable.

We showed that at much lower doping concentrations, and in the presence of a weak magnetic field, the V-doped \WSe~can show a partial ferromagnetic order, which has been observed in experiments~\cite{V_dope,Zhang2020}.
Unfortunately, such presence of magnetic order is not robust, and a uniform ferromagnetic order may not exist at such low doping.
Moreover, the energetic favorability of clustering, the lowering of magnetic exchange due to clustering, and the presence of some level of ferromagnetic order at low doping due to very high magnetic exchange in unclustered systems combined can explain the experimental observation of loss of magnetic order at higher doping densities in V-doped \WSe.

Given that the clustering of dopants is a local effect, clustering could explain the inconsistencies observed in various experiments related to doped TMDs.
Although high critical temperature can be realized in \WSe, it will be subject to experimental conditions, and realizing high Curie temperature in doped TMDs is only possible if dopant clustering can be avoided.
\new{In this work, we have limited ourselves to the clustering of dopants, however, a promising avenue to further understand the variability of magnetic order in doped TMDs could be to study the effect of vacancies.}

%Finally, we would like to mention that there are some recent experimental reports on FM order in \WSe~\cite{V_dope} and other TMDs doped with vanadium~\cite{sample31,sample32}, the effects mentioned in this work such as the short-range domain formation can be observed experimentally.

\section{Methodology}\label{s:method} 

All the ab-initio DFT calculations reported in this work were performed using the Vienna ab-initio simulation package~\cite{sample3,sample5}.
The ground state self-consistent field (scf) calculations were performed using a projector-augmented wave (PAW) potential~\cite{sample3} with a generalized-gradient approximation as proposed by the Perdew-Burke-Ernzerhof (PBE)~\cite{sample4}.
We have used a kinetic energy cut-off of 450 eV for our DFT calculations.
{All scf calculations have been performed with spin-orbit coupling (SOC) to take into account the long-range interaction~\cite{Duong2019}.}
The Brillouin zones were sampled using a $\Gamma$-centred $k$-point mesh of size $5\times5\times1$ for $4\times4\times1$, and $3\times3\times1$ for $5\times5\times1$ and $7\times7\times1$ supercells.
The $\mathrm{WSe_2}$ supercells doped with transition-metals were relaxed till the force on each of the ions was less than $0.01 \mathrm{eV}/\mathrm{\AA}$.
The energy convergence criterion for the subsequent scf calculations was set to $10^{-4}\,\mathrm{eV}$.

We have used the Hubbard $U$ model within DFT$+U$~\cite{sample6} to take into account the electron-electron interaction in the $d$ orbital of the magnetic transition-metal atoms.
We use the linear response method~\cite{sample2} to determine the Hubbard $U$ parameter.
The $U$ value we obtain from the linear response is $4$ eV, 4.5 eV, and $5.5$ eV for V, Fe, and Mn respectively.

For obtaining the parameters of Eq.~(\ref{e:J}), we put two dopant atoms in a supercell of \WSe~at various positions (near, next, and next-next neighbor) and calculate the total energy of various magnetic configurations.
We use supercells of sizes $4\times4\times1$, $5\times5\times1$, and $7\times7\times1$ to calculate the total energies of various magnetic configurations.
We also put compressive and tensile stress in $4\times4\times1$ supercells, calculate the total energies and use them along with the pristine lattices to obtain the $J$ parameters as a function of distance.

For the MC simulations, to ensure that we capture the effect of configurational entropy, for each doping concentration, we take 25 randomly doped configurations of a $30\times30\times1$ supercell of \WSe~and to calculate $T_{\rm C}$ for each of the configurations.

For studying the effect of clustering, we dope the above-mentioned supercells with more than two dopants and calculate the total energy of their magnetic configurations, and use our computational model to obtain the $J$ parameters.
However, clustering causes the $J$ parameters to change significantly and it is impossible to obtain unique $J$ parameters that work both for the unclustered and the clustered configurations.
To overcome this limitation, we use statistical weighting to obtain parameters.
We divide the clustered and the unclustered structures into two groups and write the objective function as,
\begin{equation}
W
= \sqrt{\sum_k\alpha_k\sum_i|E_{k,i}-E^{\mathrm{NM}}_k-H_{k,i}-H_{k,i}^{\mathrm{self}}|^2}.
\label{e:objectivefunction}
\end{equation}

Here, we define a statistical weight parameter $\alpha_k$ for each geometric configuration ($k^{\rm th}$) and tune this parameter depending on the clustering we want to include.
For example, for obtaining parameters suitable for 10\% clustering, we define $\alpha_k=0.1$ for $k\in{\rm clustered}$ and $\alpha_k=0.9$ for $k\in{\rm unclustered}$.

\new{To quantify the stability of clusters, we performed lattice-switch Monte-Carlo (MC) simulations~\cite{LMC}, where we parameterized the energy barriers between different lattice configurations using the formation energy provided in Fig. 2 (a).
For lattice-switch MC, we move a dopant atom from its initial position to a different position and estimate the total energy of the system using the above formula for both configurations (e.g., $l$ and $m$).
We next employ the Metropolis algorithm to determine if the new dopant configuration can be chosen or rejected with a probability $\exp{-E/k_{\rm B} T}$, where $\Delta E=E_l-E_m$, $k_{\rm B}$ is the Boltzmann constant and $T$ is the temperature. 
To quantify the clustering of dopants, we calculate the average dopant distance as a function of temperature (refer to the supplementary document for more details)~\cite{SI}).}

\section{Acknowledgements}
The project or effort depicted was or is sponsored by the Department of Defense, Defense Threat Reduction Agency.
The content of the information does not necessarily reflect the position or the policy of the federal government, and no official endorsement should be 
inferred.

\new{This work at the University of Texas at Dallas is supported by Office of Naval Research (ONR) grant no. N00014-23-1-2020.}

This work was supported in part by NEWLIMITS, a center in nCORE, a Semiconductor Research Corporation (SRC) program sponsored by NIST through award number 70NANB17H041. 

This work was also supported in part by SUPREME, one of seven centers in JUMP 2.0, a Semiconductor Research Corporation (SRC) program sponsored by DARPA.

This work was supported by imec's Industrial Affiliation Program.

\section{Supporting Information}

\newer{Further information on lattice-switch Monte-Carlo method (Section S1), results for MC simulation of 14\%~doping concentration, and formation energy as a function of Se chemical potential (Section S2).}~\cite{SI}

\section*{References}
\bibliographystyle{achemso}
\bibliography{bib}

\captionsetup{labelformat=empty}

\begin{figure*}[th]
\captionsetup{labelformat=empty,listformat=empty}
  \centering
%  \subfigure[]{\includegraphics[width=1.8\columnwidth]{clustr.pdf}}\\
  {\includegraphics[width=\columnwidth]{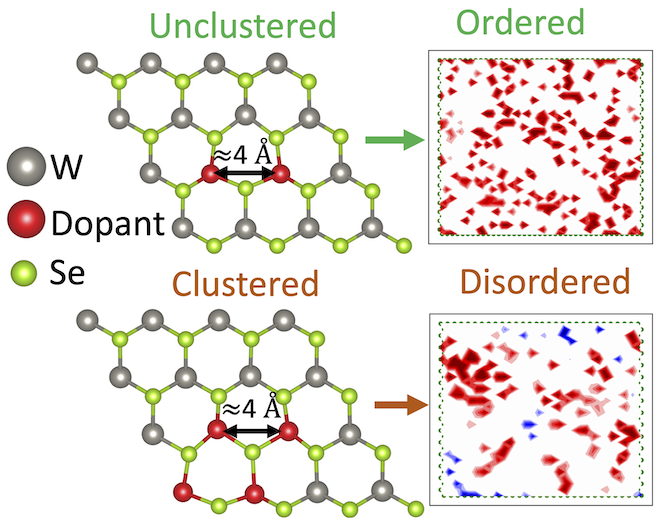}}
  \caption*{Table of content graphics.} 
\end{figure*}

\end{document}